\documentclass[aps,floatfix,superscriptaddress,preprintnumbers,amsmath,amssymb,showpacs,showkeys]{revtex4}
\usepackage{amssymb}
\usepackage{amsmath}
\usepackage{amsfonts}
\usepackage{epsfig}
\usepackage{graphicx}
\usepackage{tabularx}
\usepackage{color}
\newcommand{\seq}{\begin{subequations}}
\newcommand{\sen}{\end{subequatons}}
\newcommand{\eq}{\begin{eqnarray}}
\newcommand{\en}{\end{eqnarray}}

\def\shiftdown#1{#1\llap{\lower.04ex\hbox{#1}}}

\begin{document}

\title{Deuteron electromagnetic structure functions \\
and polarization properties in soft-wall AdS/QCD} 
                                                                            
\author{Thomas Gutsche}
\affiliation{Institut f\"ur Theoretische Physik,
Universit\"at T\"ubingen, \\
Kepler Center for Astro and Particle Physics,
Auf der Morgenstelle 14, D-72076 T\"ubingen, Germany}
\author{Valery E. Lyubovitskij}
\affiliation{Institut f\"ur Theoretische Physik,
Universit\"at T\"ubingen, \\
Kepler Center for Astro and Particle Physics,
Auf der Morgenstelle 14, D-72076 T\"ubingen, Germany}
\affiliation{Department of Physics, Tomsk State University,
634050 Tomsk, Russia}
\affiliation{Laboratory of Particle Physics, 
Mathematical Physics Department,
Tomsk Polytechnic University,
Lenin Avenue 30, 634050 Tomsk, Russia}
\author{Ivan Schmidt}
\affiliation{Departamento de F\'\i sica y Centro Cient\'\i
fico Tecnol\'ogico de Valpara\'\i so (CCTVal), Universidad T\'ecnica
Federico Santa Mar\'\i a, Casilla 110-V, Valpara\'\i so, Chile}

\vspace*{.2cm}

\date{\today}

\begin{abstract}

We apply a soft-wall AdS/QCD approach to the description of 
deuteron structure functions and tensor polarized properties. 
Present work is a completion of our previous study on 
electromagnetic form factors. By the appropriate choice 
of two couplings in the effective action we are able to reproduce 
both the form factors and structure functions in full consistency with 
model-independent constraints set by perturbative QCD. 
Our framework is based on a five-dimensional action in 
AdS space formulated in terms of vector fields dual to the deuteron 
and the electromagnetic fields. 
The AdS fields depend on four Minkowski and one holographic coordinate $z$. 
The scale dependence of the form factors and structure functions is consistent 
with quark counting rules implying the $1/Q^{10}$ behavior of the charge form 
factor $G_C(Q^2)$, structure functions $A(Q^2)$ and $B(Q^2)$, and 
the $1/Q^{12}$ behavior of the magnetic 
$G_M(Q^2)$ and quadrupole $G_Q(Q^2)$ form factors. 
 
\end{abstract}

\vspace*{.5cm}
\date{\today}



\maketitle

\vspace*{.5cm}

\section{Introduction}

The study of electron-deuteron scattering gives a unique insight 
into the structure of the deuteron. 
The structure information is contained in the
gauge-invariant matrix element describing the interaction of 
the deuteron with the electromagnetic field, which 
reads as 
\eq\label{M_inv}
M_{\rm inv}^\mu(p,p') &=& -
\biggl( G_1(Q^2) \epsilon^\ast(p') \cdot \epsilon(p) -
\frac{G_3(Q^2)}{2M_d^2} \, \epsilon^\ast(p') \cdot q \,
\epsilon(p) \cdot q \biggr) \, (p+p')^\mu \nonumber\\
&-& G_2(Q^2) \,
\biggl( \epsilon^\mu(p) \, \epsilon^{\ast}(p') \cdot q
- \epsilon^{\ast\mu}(p') \, \epsilon(p) \cdot q \biggr)
\en
where  $\epsilon$($\epsilon^\ast$) and
$p(p^\prime)$ are the polarization and four-momentum of the initial (final)
deuteron, $q=p^\prime - p$ is the momentum transfer and $M_d$ is the 
deuteron mass. 

The three electromagnetic (EM) form factors $G_{1,2,3}$ of the deuteron
are related to the charge $G_C$, quadrupole $G_Q$ and magnetic $G_M$
form factors by
\eq
G_C = G_1+\frac{2}{3}\tau_d G_Q\,, \hspace*{.25cm}
G_M \ = \ G_2 \,,            \hspace*{.25cm}
G_Q = G_1-G_2+(1+\tau_d)G_3,   \hspace*{.25cm}
\tau_d=\frac{Q^2}{4M_d^2} \,.
\en
The form factors are normalized at $Q^2 = 0$ as 
\eq
G_C(0)=1\,, \ \ \
G_Q(0)=M_d^2{\cal Q}_d=25.83\,, \ \ \
G_M(0)=\frac{M_d}{M_N}\mu_d=1.714 \, ,
\en
where $M_N$ is the nucleon mass, ${\cal Q}_d = 7.3424$ GeV$^{-2}$ and 
$\mu_d = 0.8574$ are the quadrupole and magnetic moments of the deuteron.

The set of the three form factors ($G_C(Q^2)$, $G_M(Q^2)$, $G_Q(Q^2)$) 
and their combinations --- structure functions 
\eq 
A(Q^2) = G_C^2(Q^2) + \frac{2}{3} \tau_d G_M^2(Q^2)
+ \frac{8}{9} \tau_d^2 G_Q^2(Q^2)\,, \quad
B(Q^2) = \frac{4}{3} \tau_d (1 + \tau_d) G_M^2(Q^2) \,.
\en
are the main observables defining the electromagnetic structure
of the deuteron (for a recent review on the experimental 
and theoretical progress 
see Ref.~\cite{Holt:2012gg,Marcucci:2015rca,Gutsche:2015xva,Dong:2008mt}). 
In Ref.~\cite{Brodsky:1983vf,Carlson:1984wr,Brodsky:1992px} model-independent 
relations between the deuteron form factors and their individual scalings 
imposed by perturbative QCD at large Euclidean values of $Q^2$ 
have been derived. In particular, in Ref.~\cite{Brodsky:1983vf}  
it was shown that $\sqrt{A(Q^2)}$ 
can be factorized in terms of the nucleon form factor
$F_N(Q^2/4)$ and the so-called ``reduced'' nuclear form factor
$f_d(Q^2)$ as
\eq 
\sqrt{A(Q^2)} = f_d(Q^2) \, F_N^2(Q^2/4) \,.
\en 
$\sqrt{A(Q^2)}$ scales as $1/Q^{10}$ in agreement with quark counting rules, 
since the deuteron has six constituent quarks. 
Ref.~\cite{Carlson:1984wr} contains a derivation of the high $Q^2$ 
QCD prediction for the following relation between the charge and quadrupole 
form factors 
\eq 
G_C(Q^2) = \frac{2}{3} \tau_d \, G_Q(Q^2) \,. 
\en 
In Ref.~\cite{Brodsky:1992px} this relation has been extended by including 
the next-to-leading term in the form 
\eq 
G_C(Q^2) = \biggl(\frac{2}{3} \tau_d - 1\biggr) 
\, G_Q(Q^2) \,. 
\en 
In addition it was shown that the three deuteron form factors 
and structure functions scale at large $Q^2$ as 
\eq 
& &G_C(Q^2) \, \colon \, G_M(Q^2) \, \colon \, G_Q(Q^2) =  
\biggl(\frac{2}{3} \tau_d - 1\biggr) \, \colon \, 2 \, \colon \, - 1 \,, 
\nonumber\\
& &B(Q^2) \, \colon \, A(Q^2) \, \colon \, 
G_C^2(Q^2) = 4 \, \colon \, 1 \, \colon \, \frac{1}{3} 
\,. 
\en 
Using these results of 
Refs.~\cite{Brodsky:1983vf,Carlson:1984wr,Brodsky:1992px} 
one concludes that the individual deuteron form factors and 
structure functions scale at large $Q^2$ as 
\eq\label{scalings}
& &\sqrt{A(Q^2)} \sim  \sqrt{B(Q^2)} \sim G_C(Q^2) \sim \frac{1}{Q^{10}}\,, 
\nonumber\\ 
& &G_M(Q^2) \sim G_Q(Q^2) \sim \frac{1}{Q^{12}}\,. 
\en  
The cross section for elastic electron-deuteron scattering 
is defined by the Rosenbluth formula 
\eq 
\frac{d\sigma}{d\Omega} &=& \sigma_M \, S(Q^2)\,, \nonumber\\
S(Q^2) &=& 
A(Q^2) + B(Q^2) \tan^2\frac{\theta}{2}  \,, 
\en 
where $\sigma_M$ is the Mott cross section and $\theta$ is the electron 
scattering angle.  

Other important characteristics of the deuteron are the 
deuteron tensor-polarized observables (tensor analyzing powers) 
$T_{20}$, $T_{21}$ and $T_{22}$, which are expressed in terms of 
the three form factors ($G_C(Q^2)$, $G_M(Q^2)$, $G_Q(Q^2)$) 
and scattering angle $\theta$ as 
\eq
T_{20}(Q^2,\theta) &=& - \frac{1}{S(Q^2) \sqrt{2}} \,
\biggl[
   \frac{8}{3} \, \tau_d   \, G_C(Q^2) \, G_Q(Q^2) \nonumber\\
&+&\frac{8}{9} \, \tau_d^2 \, G_Q^2(Q^2)
 + \frac{\tau_d}{3} \Big(1 + 2 (1 + \tau_d) \tan^2\frac{\theta}{2}\Big)
 \, G_M^2(Q^2)
\biggr] \,, \\
T_{21}(Q^2,\theta) &=& - \frac{2}{S(Q^2) \sqrt{3}} \,
\sqrt{\tau_d^3 \Big( 1 + \tau_d \sin^2\frac{\theta}{2} \Big)} \,
\frac{G_M(Q^2) G_Q(Q^2)}{\cos\frac{\theta}{2}} \,,\\
T_{22}(Q^2,\theta) &=& - \frac{1}{2 S(Q^2) \sqrt{3}} \,
\tau_d \, G_M^2(Q^2) \,.
\en
As shown in Ref.~\cite{Zhang:2011zu} it is useful 
to define the quantity 
\eq
\tilde T_{20}(Q^2) = \frac{1}{1 - \delta} \,
\biggl( T_{20}(Q^2) + \frac{\delta}{2 \sqrt{2}}
\biggr)
\en
with the use of the factor 
\eq
\delta = \frac{B(Q^2)}{S(Q^2)} \,
\biggl[ \frac{1}{2 (1 + \tau_d)} + \tan^2\frac{\theta}{2}  \biggr]
\en
in order to eliminate the dependence of $T_{20}$ on $G_M$ and $\theta$: 
\eq
\tilde T_{20}(Q^2) = - \frac{\tau_d}{\sqrt{2}} \,
\frac{3 \beta(Q^2) + \tau_d}{\frac{9}{8} \beta^2(Q^2) + \tau_d^2}\,,
\en
where $\beta(Q^2) = G_C(Q^2)/G_Q(Q^2)$ is the ratio of $G_C$ and $G_Q$ 
form factors. By further elimination of the leading $Q^2$ dependence 
one can introduce the reduced quantity 
\eq
\tilde T_{20R}(Q^2) &=& - \frac{3}{Q^2 Q_d \sqrt{2}} \,
\tilde T_{20}(Q^2) \nonumber\\
&=& \frac{G_Q(Q^2)}{G_Q(0)} \,
\frac{G_C(Q^2) + \frac{\tau_d}{3} G_Q(Q^2)}
{G_C^2(Q^2) + \frac{8}{9} \tau_d^2 G_Q^2(Q^2)} \,,
\en
which is normalized as $\tilde T_{20R}(0) = 1$ for $Q^2 = 0$. 

Using the power scalings of the deuteron form factors~(\ref{scalings}) 
one can derive large $Q^2$ scalings for 
the tensor-polarized observables~\cite{Brodsky:1992px} as 
\eq
& &T_{20}(Q^2) \sim - \sqrt{2} \,
\frac{1 + \tan^2\frac{\theta}{2}}{1 + 4 \tan^2\frac{\theta}{2}}  
\sim {\cal O}(1)
\,,\nonumber\\
& &T_{21}(Q^2) \sim   \sqrt{3} \,
\frac{\tan\frac{\theta}{2}}{1 + 4 \tan^2\frac{\theta}{2}} \sim {\cal O}(1)
\,,\nonumber\\
& &T_{22}(Q^2) \sim   -\frac{\sqrt{3}}{2 \tau_d} \,
\frac{1}{1 + 4 \tan^2\frac{\theta}{2}} \sim {\cal O}(1/Q^2)
\,,\nonumber\\
& &\tilde T_{20}(Q^2) \sim   - \sqrt{2} \sim {\cal O}(1)
\,,\nonumber\\
& &\tilde T_{20R}(Q^2) \sim \frac{3}{4 \tau_d G_Q(0)} \sim {\cal O}(1/Q^2) \,.
\en

In the present paper we further improve the calculation 
on the deuteron form factors 
in a soft-wall AdS/QCD approach~\cite{AdSQCD_int1,AdSQCD_int3,%
AdSQCD1,AdSQCD2}, started in Ref.~\cite{Gutsche:2015xva}.  
We present in addition the results for the structure functions and the 
tensor-polarized variables. Our approach is based on gauge/gravity duality 
and is constructed as a holographic dual to pQCD.
The main advantage of our framework is that it gives a description of
the deuteron electromagnetic form factors in the full $Q^2$ regime 
and guarantees the correct power scaling of form factors and structure 
functions at large $Q^2$. Note that this important property of 
approaches based on gauge/string duality originally was originally found  
in Ref.~\cite{Polchinski:2001tt} and was later confirmed in 
Refs.~\cite{AdSQCD_int3,AdSQCD1,AdSQCD2,Gutsche:2015xva}.   

\section{Formalism and numerical results}

The formalism is based on an effective action given in terms of 
the AdS fields $d^M(x,z)$ and $V^M(x,z)$. These fields are
dual to the Fock component contributing to the deuteron with 
twist $\tau = 6$ and the electromagnetic field, respectively. 
The action is given by
\eq\label{Eff_action}
S &=& \int d^4xdz \, \sqrt{g} \, e^{-\varphi(z)} \,
\biggl[ - \frac{1}{4} F_{MN} F^{MN}
- D^M d^\dagger_{N} D_M d^N
- i c_2(z) F^{MN} d^\dagger_{M} d_{N}\nonumber\\
&+& \frac{c_3(z)}{4M_d^2} \, e^{2A(z)} \, \partial^M F^{NK}
\biggl( iD_K d^\dagger_{M} d_{N}
- d^\dagger_{M} i D_K d_{N} + \mathrm{H.c.}
\biggr) +
d^\dagger_{M} \, \Big(\mu^2 +  U(z) \Big) \, d^M
\biggr]  \,,
\en
where 
$g = |{\rm det g_{MN}}| = e^{10 A(z)}$ and 
$A(z) = \log(R/z)$; $F^{MN}$ is the stress tensor of vector field $V^M$; 
$D^M$ is the covariant derivative including $V^M$; 
$\mu^2 R^2 = (\Delta - 1) (\Delta - 3)$ is the five-dimensional mass;
$R$ is the AdS radius, $\varphi(z) = \kappa^2 z^2$ is the
background dilaton field and $\kappa$ is the scale parameter; 
$\Delta = \tau + 1$ is the dimension of the $d^M(x,z)$ field; 
$M_d$ is the deuteron mass and 
$U(z) = U_0 \varphi(z)/R^2$ is the confinement potential, 
where the constant $U_0$ is fixed by the value of the deuteron mass. We work 
in the axial gauge for both vector fields $d^z(x,z) = 0$ and $V^z(x,z) = 0$. 
The $z$-dependent couplings $c_2(z)$ and $c_3(z)$ are constrained 
by the normalization and the large $Q^2$ scaling of 
the deuteron electromagnetic form factors (see discussion below). 

This updated action is an extension of the deuteron action suggested 
in Ref.~\cite{Gutsche:2015xva}. 
The main difference is that now we incorporate the $z$-dependence 
of the AdS/QCD couplings $c_2$ and $c_3$:
\eq
c_2(z) &=&    e^{- \beta\varphi(z)}  \, \Big[ c_2^{(1)} \,
        + \,  c_2^{(2)} e^{\alpha_2 \log\varphi(z)}\Big] \,,
\nonumber\\
c_3(z) &=& c_3  \, e^{- \beta\varphi(z) + \alpha_3 \log\varphi(z)}\,,
\en
where the couplings $c_2^{(1)}$, $c_2^{(2)}$ and
$c_3$ are fixed from the normalization
of $G_M(Q^2)$, $G_Q(Q^2)$ and the
asymptotics of $G_M(Q^2)$ at large $Q^2$:
\eq
c_2^{(1)} &=& 12 \int\limits_0^1 dx \frac{(1-x)^5}{[1 + \beta (1-x)]^7} \,,
\nonumber\\
c_2^{(2)} &=& G_2(0) - c_2^{(1)} = \frac{M_d}{M_N} \, \mu_d  \, - \, 
c_2^{(1)} \,, \nonumber\\
c_3       &=& G_3(0) = M_d^2 Q_d - 1 + \frac{M_d}{M_N} \, \mu_d \,.
\en
The parameters $\alpha_2, \alpha_3$ and
$\beta$ should satisfy the conditions
\eq\label{alpha_beta_constraints}
\alpha_2 > 0\,, \quad
\alpha_3 > 1\,, \quad
\beta \ge 0
\en 
to guarantee the scaling behavior~(\ref{scalings})
of the deuteron form factors and structure function.  
Their numerical values are optimized in a best fit to data:
\eq\label{alpha_beta_num} 
\alpha_2 = 0.25\,, \ \alpha_3 = 1.1\,, \
\beta = 1.2  \,.
\en  

Our calculation of deuteron form factors proceeds in several steps. 
First, we perform a Kaluza-Klein (KK) decomposition of
the vector AdS field dual to the deuteron
$d^\mu(x,z)
= \exp[(\varphi(z)-A(z))/2] \,
\sum\limits_n d^\mu_{n}(x) \Phi_{n}(z)$\,, 
where $d^\mu_{n}(x)$ is the tower of KK fields dual to
the deuteron fields with radial quantum number $n$ and twist-dimension
$\tau = 6$, and $\Phi_{n}(z)$ are their bulk profiles in the fifth 
direction of the AdS space. 

Second, we derive a Schr\"odinger-type equation of motion (EOM) for
the profile $\Phi_{n}(z)$ 
\eq
\biggl[ - \frac{d^2}{dz^2} + \frac{99}{4z^2}
+ \kappa^4 z^2 + \kappa^2 U_0 \biggr] \Phi_{n}(z) =
M_{d,n}^2 \Phi_{n}(z) \,, 
\en
which is solved analytically 
\eq
\Phi_{n}(z) = \sqrt{\frac{2n!}{(n+5)!}}  \,
\kappa^{6} \, z^{11/2} \, e^{-\kappa^2 z^2/2}
\, L_n^{5}(\kappa^2z^2)\,, \ \ \ 
M_{d,n}^2
= 4\kappa^2 \biggl[ n + 3 + \frac{U_0}{4}\biggr]\,,
\en
where $L_n^m(x)$ are the generalized Laguerre polynomials.
Restricting our considerations to the ground state $n=0$ we get
$M_d = 2 \kappa \, \sqrt{3 + \frac{U_0}{4}}$.
Using the central experimental value for the deuteron mass $M_d = 1.875613$ GeV
and setting the  parameter $\kappa = 190$ MeV 
(obtained from a fit to the data of the
electromagnetic deuteron form factors), we fix $U_0 = 85.4494$.
Note that the deuteron scale parameter is two times smaller than 
the analogous parameter $\kappa_N \simeq 380$~MeV entering in the
description of the nucleon - mass and electromagnetic 
form factors~\cite{AdSQCD1,AdSQCD2}.  
The difference between the nucleon and deuteron scale parameters
can be related to the change of size of the hadronic systems -
the deuteron as a two-nucleon bound state is about
2 times larger than the nucleon.

Third, we perform a Fourier transformation of the vector field $V(x,z)$ 
with respect to the Minkowski coordinate 
\eq\label{V_Fourier}
V_\mu(x,z) = \int \frac{d^4q}{(2\pi)^4} e^{-iqx} V_\mu(q) V(q,z)
\en 
and derive an EOM for the vector bulk-to-boundary propagator $V(q,z)$ 
dual to the $q^2$-dependent electromagnetic current 
\eq
\partial_z \biggl( \frac{e^{-\varphi(z)}}{z} \,
\partial_z V(q,z)\biggr) + q^2 \frac{e^{-\varphi(z)}}{z} \, V(q,z) = 0 \,.
\en
The solution of this equation in terms of the 
gamma $\Gamma(n)$ and Tricomi $U(a,b,z)$ functions reads 
\eq
\label{VInt_q} 
V(q,z) = \Gamma\Big(1 - \frac{q^2}{4\kappa^2}\Big)
\, U\Big(-\frac{q^2}{4\kappa^2},0,\kappa^2 z^2\Big) \,. 
\en 
In the Euclidean region it is convenient to use the integral 
representation for $V(Q,z)$~\cite{Grigoryan:2007my}            
\eq 
\label{VInt}
V(Q,z) = \kappa^2 z^2 \int_0^1 \frac{dx}{(1-x)^2}
\, x^a \,
e^{- \kappa^2 z^2 \frac{x}{1-x} }\,, 
\en
where $x$ is the light-cone momentum fraction and 
$a = Q^2/(4 \kappa^2)$. 

With the present set-up the deuteron form factors 
can be easily calculated as: 
\eq
G_1(Q^2) &=&
\int\limits_0^\infty dz \Phi_0^2(z) V(Q,z) \,,\nonumber\\
G_i(Q^2) &=&
\int\limits_0^\infty dz c_i(z) \Phi_0^2(z) V(Q,z) \,, \quad i=2,3 \,.
\en
For a detailed discussion 
on the calculational technique see for example Ref.~\cite{Gutsche:2015xva}. 
The explicit expressions for the form factors are 
\eq
G_1(Q^2) &=& \frac{\Gamma(a+1) \, \Gamma(7)}{\Gamma(a+7)}\,, \nonumber\\
G_2(Q^2) &=& c_2^{(1)} \frac{I_1(Q^2)}{I_1(0)}
           + c_2^{(2)} \frac{I_2(Q^2)}{I_2(0)}\,, \nonumber\\
G_3(Q^2) &=& c_3 \frac{I_3(Q^2)}{I_3(0)} \,,
\en
where
\eq
& &I_i(Q^2) =
\frac{\Gamma(7 + \Delta_i)}{\Gamma(6)}\,
\int\limits_0^1 dx \, x^a \, \frac{(1-x)^{5 + \Delta_i}}
{(1 + \beta (1-x))^{7+\Delta_i}}
\,,\nonumber\\
& &\Delta_1 \equiv 0\,, \quad
\Delta_2 = \alpha_2\,, \quad
\Delta_3 = \alpha_3\,, \quad 
c_2^{(1)} = 2 I_1(0) \,. 
\en
Note that the scaling of our form factors, the structure functions 
and tensor-polarized observables is fully consistent 
with perturbative QCD~\cite{Brodsky:1983vf,Carlson:1984wr,Brodsky:1992px}. 
In particular, 
the structure functions $A(Q^2)$, $B(Q^2)$ and the charge 
form factor $G_C(Q^2)$ have the correct power-scaling
$1/Q^{10}$ at large $Q^2 \to \infty$, 
while the magnetic and quadrupole form factors scale as 
$1/Q^{12}$ at large $Q^2$. 
Our results for the charge $G_C(Q^2)$, quadrupole $G_Q(Q^2)$
and magnetic $G_M(Q^2)$ form factors, structure functions 
$A(Q^2)$ and $B(Q^2)$, tensor-polarized quantities $T_{20}(Q^2)$, 
$\tilde T_{20R}(Q^2)$, $T_{21}(Q^2)$ and $T_{22}(Q^2)$ 
are shown in Fig.~1-3.
Results are compared to data taken from 
Refs.~\cite{Abbott:2000ak,Holt:2012gg} and references within. 
The shaded bands correspond to values for the scale parameter $\kappa$
in the range of 150$-$250 MeV. An increase of the
parameter $\kappa$ leads to an enhancement of the form factors.
The best description of the data on the deuteron form factors is obtained
for $\kappa = 190$ MeV and is shown by the solid line.
We also determine the deuteron
charge $r_C = (-6 dG_C(Q^2)/dQ^2|_{Q^2=0})^{1/2}$ 
and 
magnetic 
$r_M = (-6 dG_M(Q^2)/dQ^2|_{Q^2=0}/G_M(0))^{1/2}$ 
radii, which for $\kappa = 190$ MeV 
are equal to $1.92$~fm and $2.26$~fm, respectively. These values 
compare well with the data of 
$r_C = 2.13 \pm 0.01$~fm and 
$r_M = 1.90 \pm 0.14$~fm. 

In the soft-wall AdS/QCD approach we are able to describe
of full $Q^2$ behavior of the electromagnetic structure of
the deuteron including the tensor-polarized observables.
One exception is the tensor analyzing power $T_{20}$ where
discrepancies between prediction and data occur for $Q$
larger than 4 fm$^{-1}$. This mismatch can be traced to the behavior 
of $G_C(Q^2)$, where the prediction overestimates the data for large Q.
The predicted behavior of $G_C(Q^2)$ 
can further be linked to the electromagnetic 
form factor $G_1(Q^2)$ contributing in the form
\eq
G_C = G_1+\frac{2}{3}\tau_d G_Q.
\en
While the prediction for $G_1 (Q^2)$ 
is positive for all values of $Q^2$, the corresponding
data of $G_C$ require a $G_1$ which should cross zero around 
$Q \sim 4$ fm$^{-1}$. 
As an illustration in Fig.~4 we present two plots for $G_C$ ---
the exact result (left panel) and where the form factor $G_1$ is dropped
(right panel), i.e.
$G_C \simeq \tilde G_C = \frac{2}{3}\tau_d G_Q$ with a  $\kappa$ of 190 MeV
and a variation in $\beta$ from 0 to 2. 
Similar plots for the magnetic, quadrupole
form factors and the structure functions are shown in Figs.~5 and 6. In all
plots an increasing $\beta$ value leads to increase of all observables 
in the $Q^2$ dependence.

From Fig.~4 it is apparent that the form factor
$G_1(Q^2)$ should cross zero around $Q \simeq 4$ fm$^{-1}$ 
in order to suppress
the contribution of the $G_Q$ form factor in $G_C$.

Finally, in Figs.~7-9 we show the sensitivity 
of all form factors and structure
functions on the variation of the parameters $\alpha_2$ and $\alpha_3$: 
$0 \le \alpha_2 \le 1\,, \ 
1 \le \alpha_3 \le 2$\,. 

\begin{figure}
\begin{center}
\includegraphics[scale=.45]{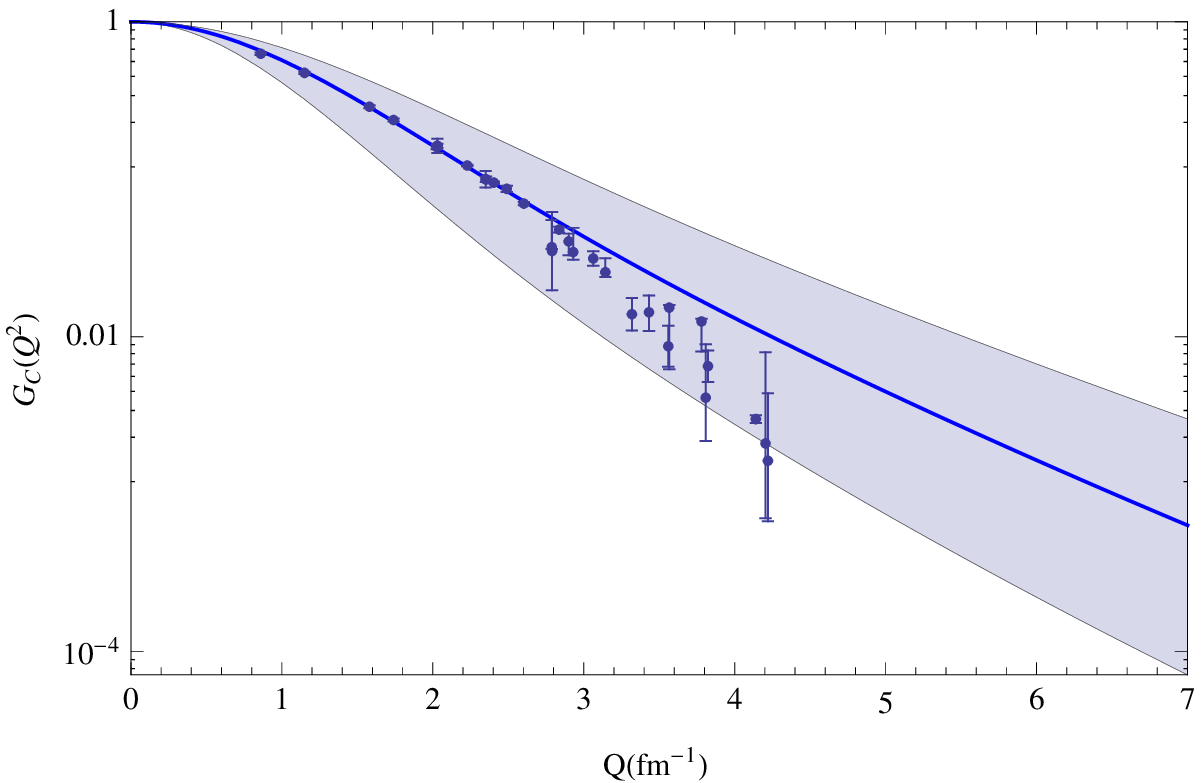} 
\includegraphics[scale=.45]{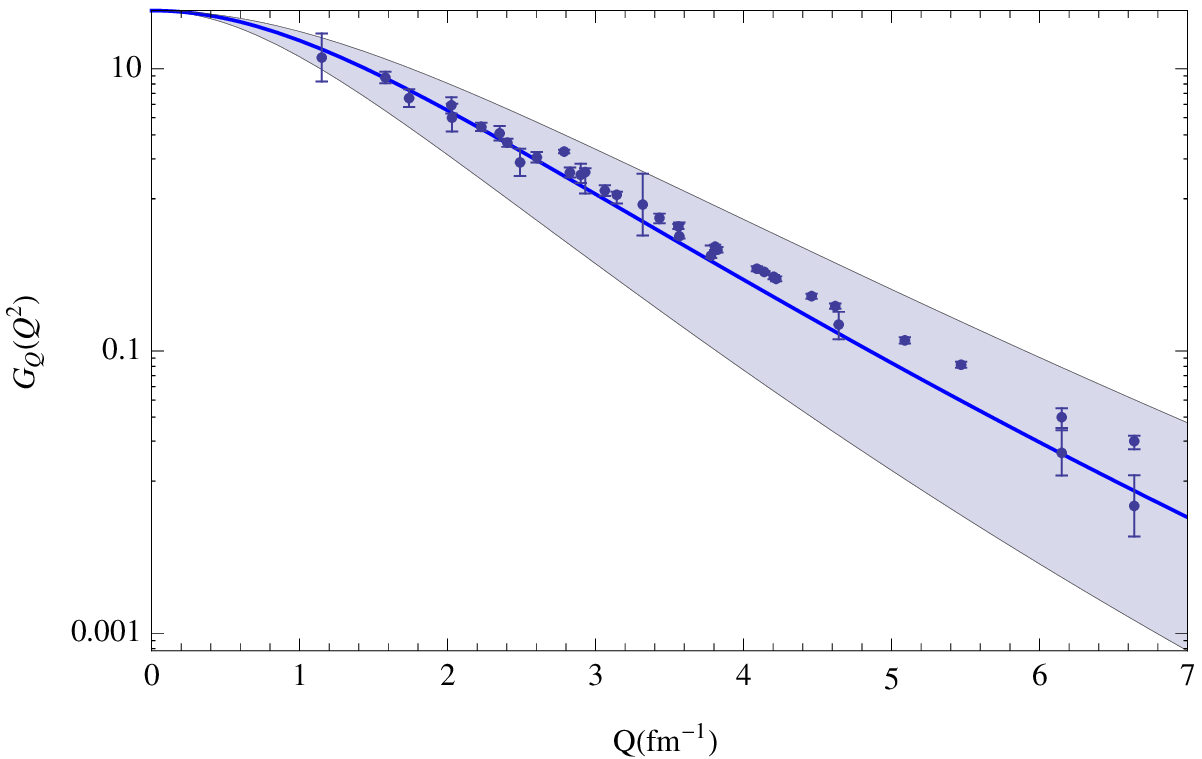} 
\includegraphics[scale=.45]{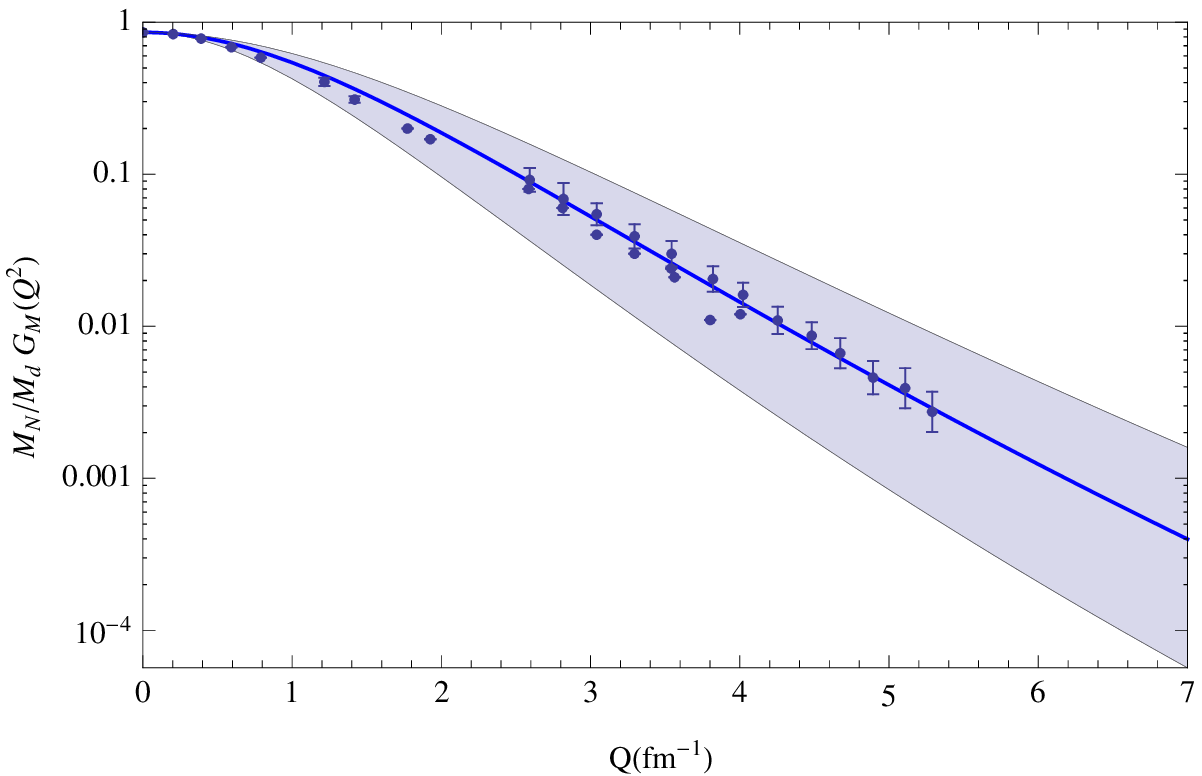}
\noindent
\caption{Charge $G_C(Q^2)$, quadrupole $G_Q(Q^2)$
and magnetic $(m_N/m_D) \, G_M(Q^2)$ deuteron form factor.}
\vspace*{.2cm}
\includegraphics[scale=.45]{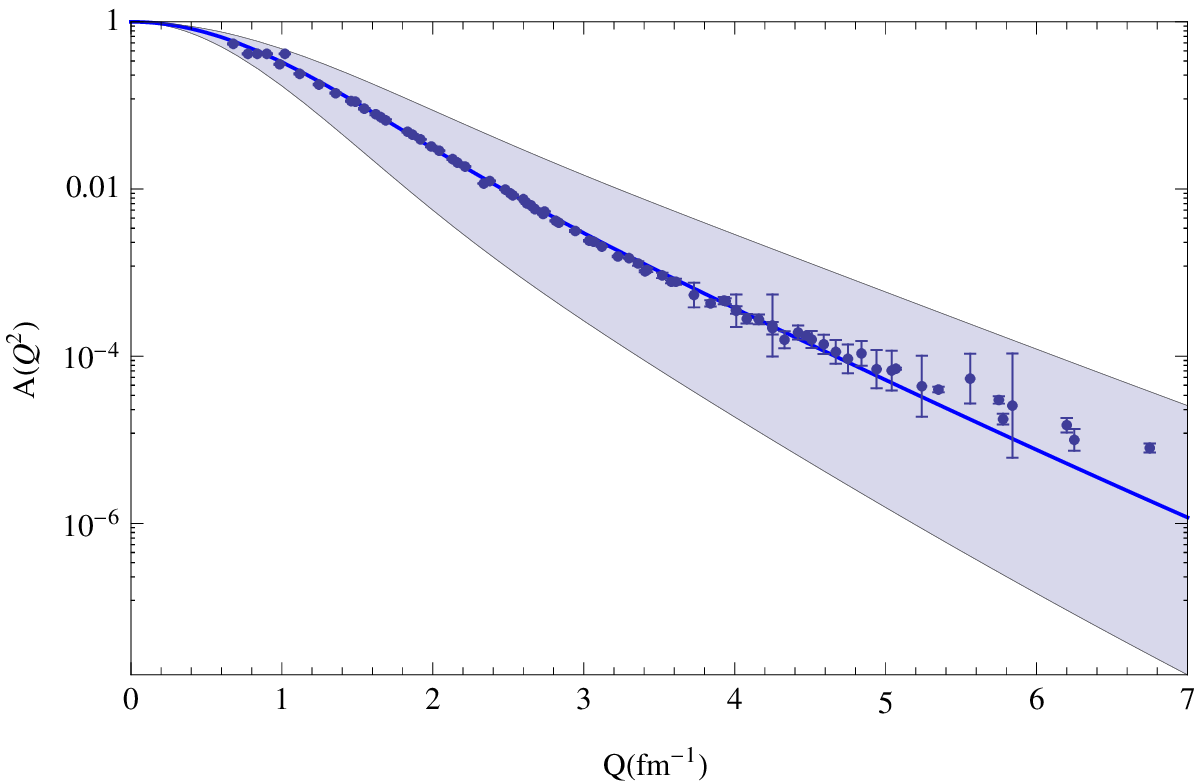} 
\includegraphics[scale=.45]{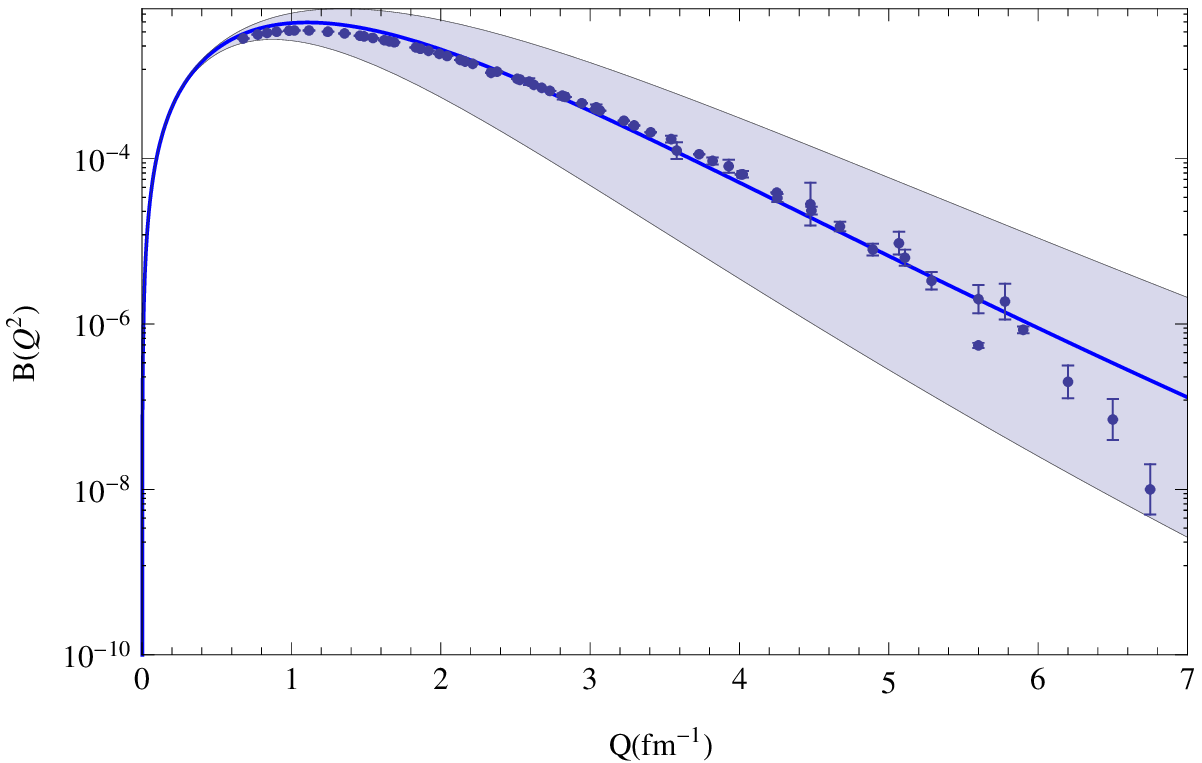} 
\caption{Deuteron structure functions $A(Q^2)$ and $B(Q^2)$.} 
\vspace*{.2cm}
\includegraphics[scale=.45]{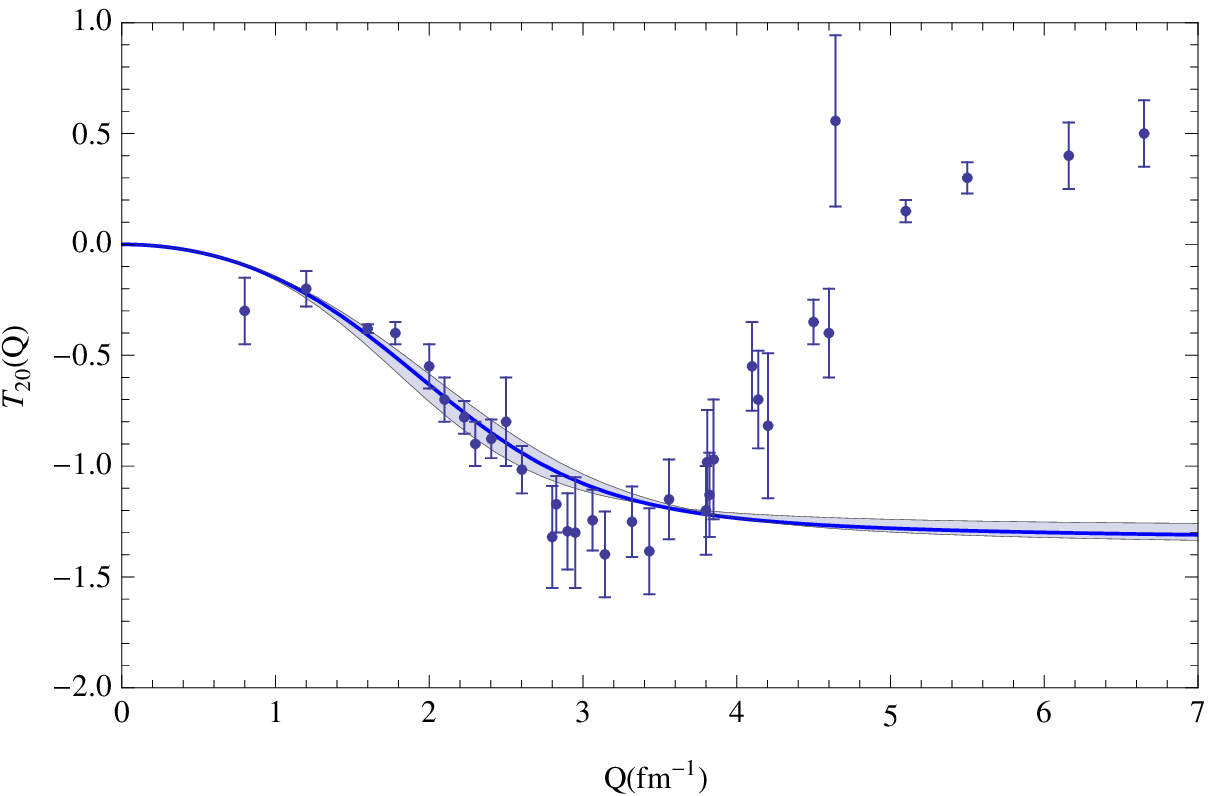} 
\includegraphics[scale=.45]{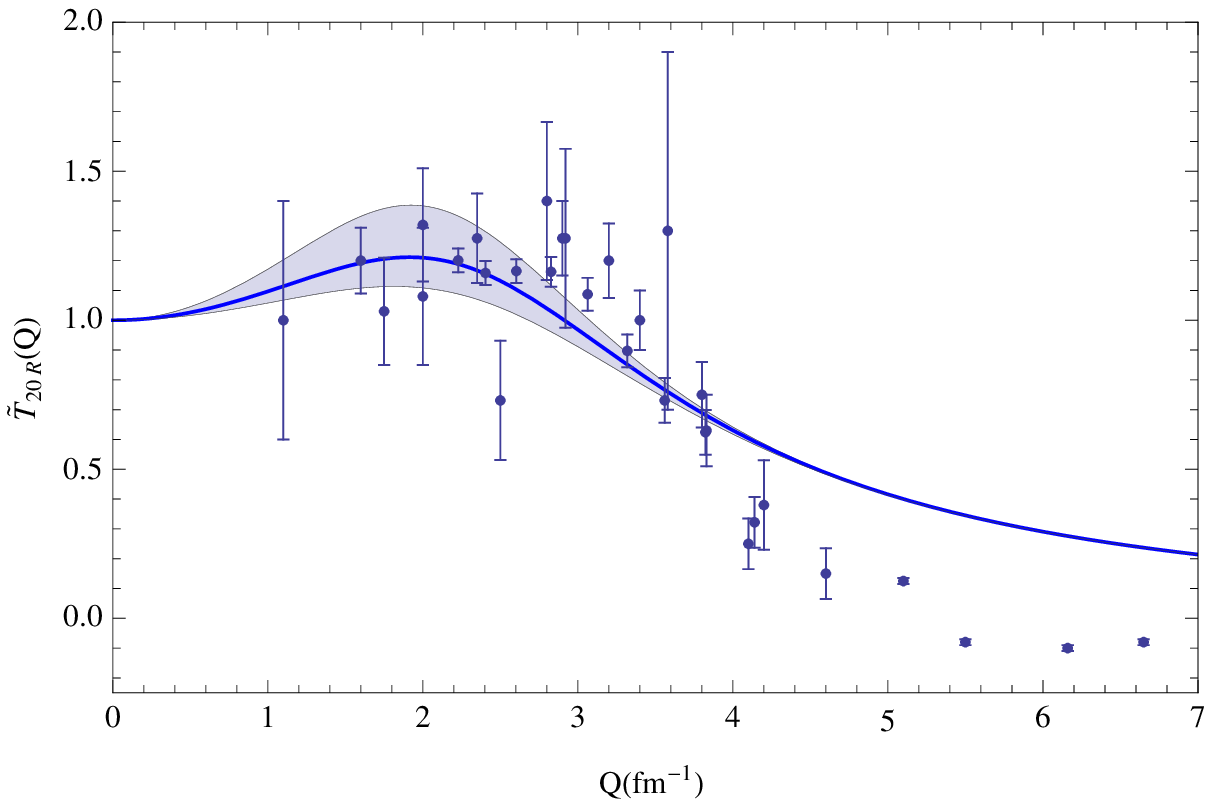}\\ 
\vspace*{.2cm}
\includegraphics[scale=.45]{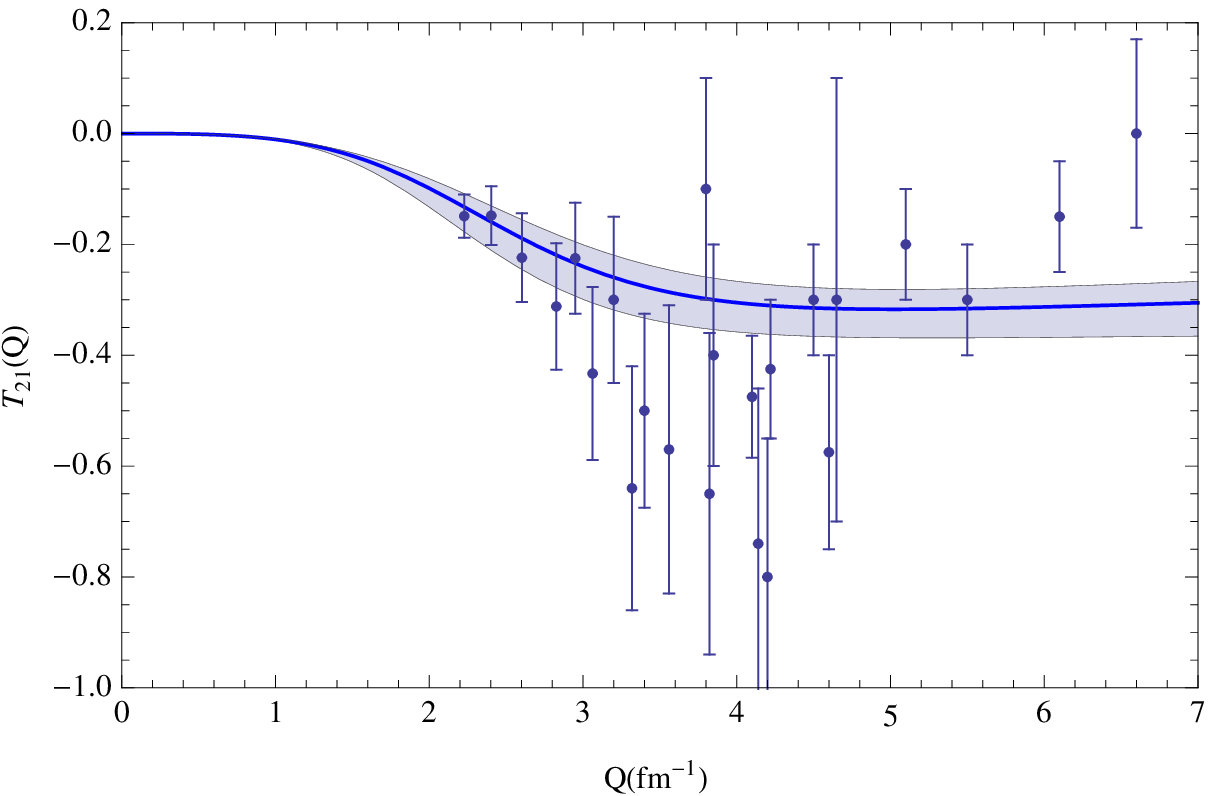} 
\includegraphics[scale=.45]{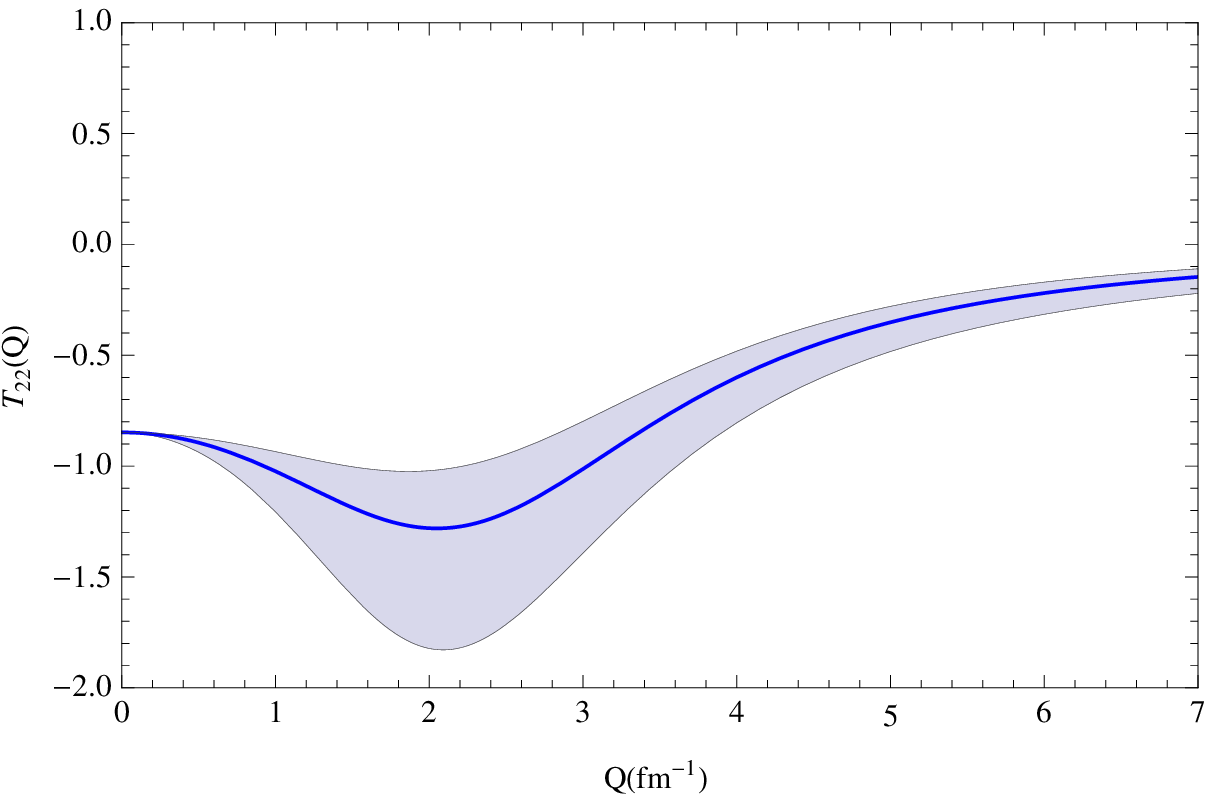} 
\caption{Deuteron tensor-polarized structures 
$T_{20}(Q^2)$, $\tilde T_{20R}(Q^2)$, $T_{21}(Q^2)$ and $T_{22}(Q^2)$.} 
\end{center}
\end{figure} 

\begin{figure}
\begin{center}
\includegraphics[scale=.45]{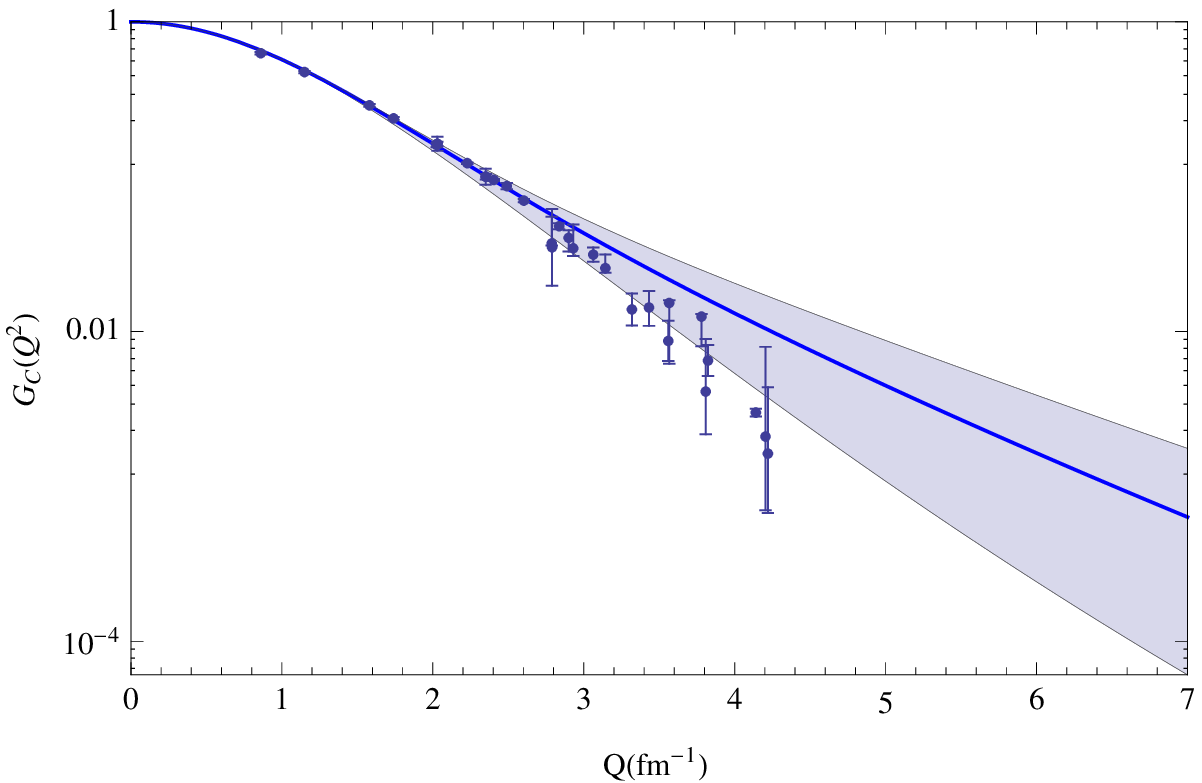}
\includegraphics[scale=.45]{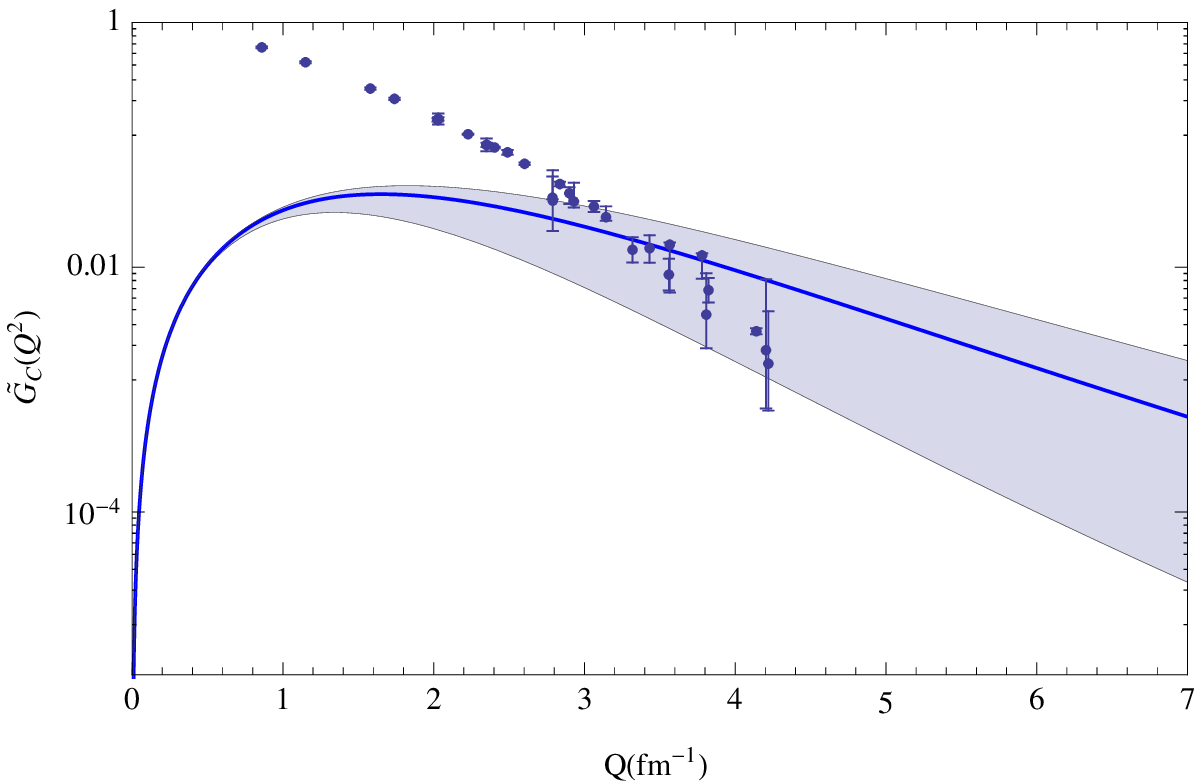}
\noindent
\caption{Results for the charge form factor $G_C(Q^2)$ with the parameters
in the range of $0 \le \beta \le 2$ 
(solid line corresponds to the $\beta = 1.2$) 
and for the fixed values $\alpha_2 = 0.25$,
$\alpha_3 = 1.1$:
exact result (left panel) and truncated result $\tilde G(Q^2)$
(right panel).} 
\vspace*{.2cm}
\includegraphics[scale=.45]{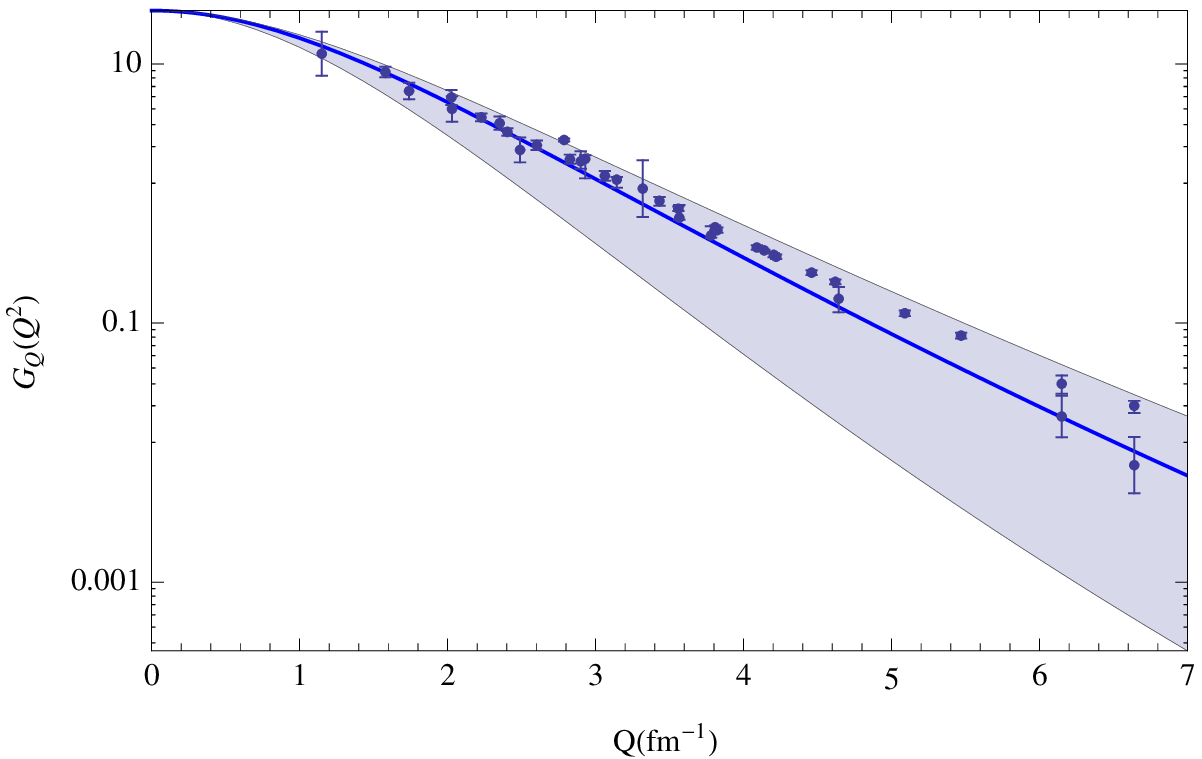}
\includegraphics[scale=.45]{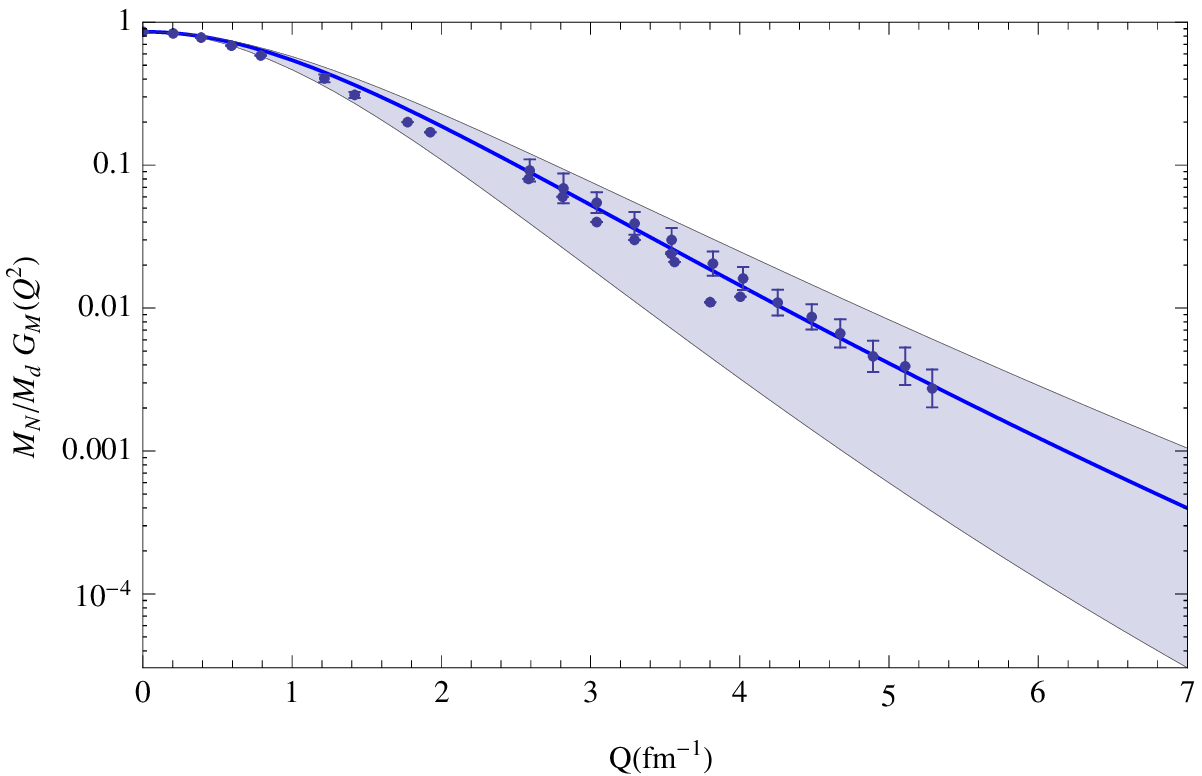}
\noindent
\caption{Quadrupole $G_Q(Q^2)$
and magnetic $(m_N/m_D) \, G_M(Q^2)$ deuteron form factors
for the parameters $0 \le \beta \le 2$ 
(solid line corresponds to the $\beta = 1.2$) 
and $\alpha_2 = 0.25$,
$\alpha_3 = 1.1$.}
\vspace*{.2cm}
\includegraphics[scale=.45]{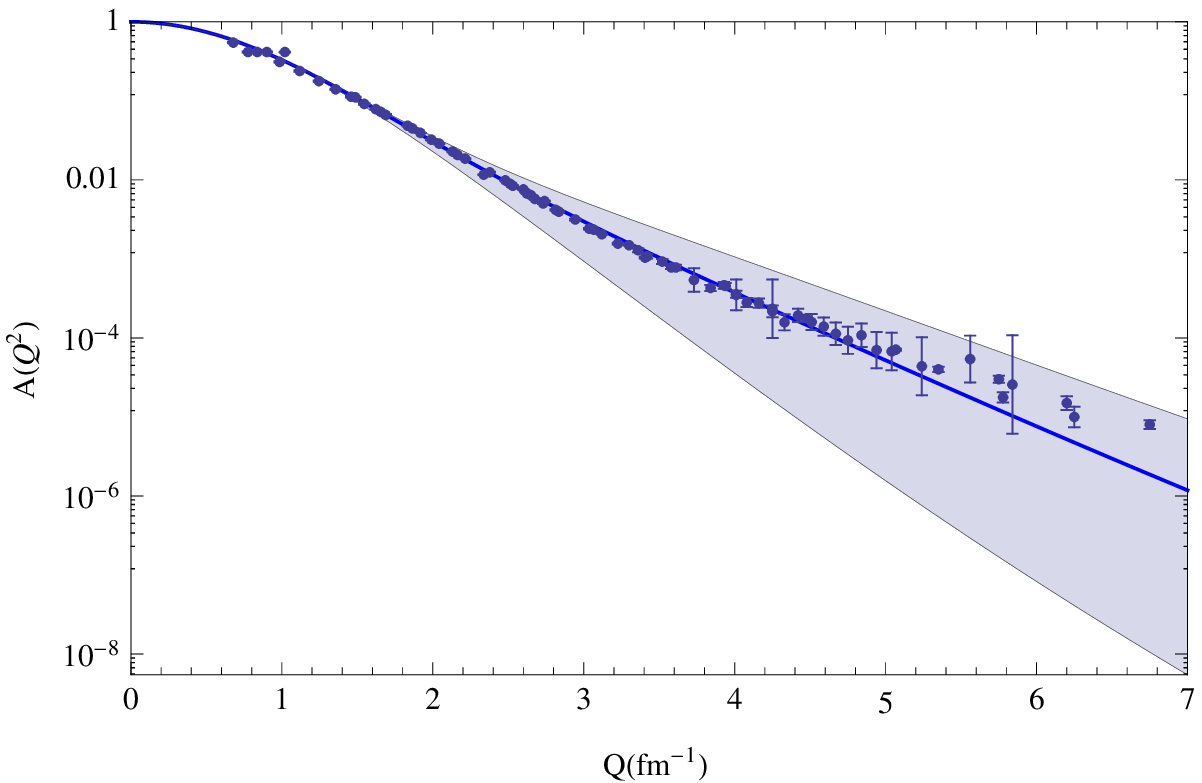}
\includegraphics[scale=.45]{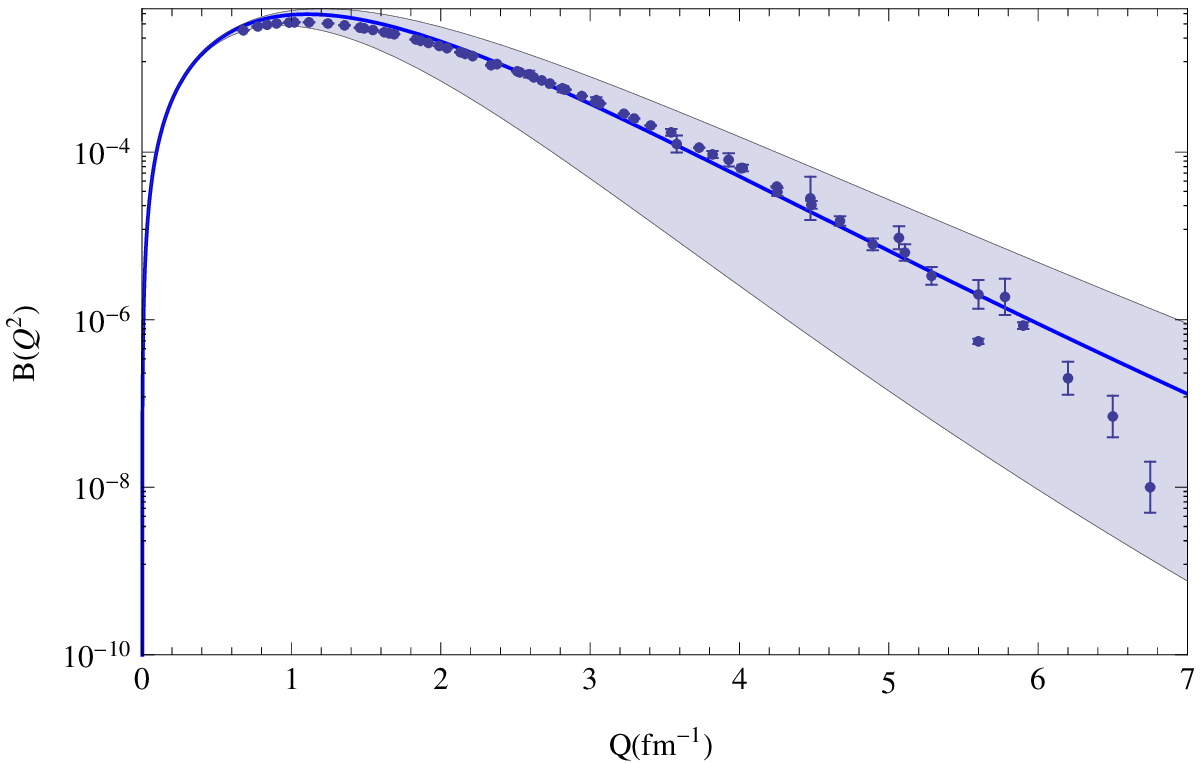}
\caption{Structure functions $A(Q^2)$ and $B(Q^2)$
for the parameters $0 \le \beta \le 2$ 
(solid line corresponds to the $\beta = 1.2$) 
and $\alpha_2 = 0.25$,
$\alpha_3 = 1.1$.}
\end{center}
\end{figure}

\begin{figure}
\begin{center}
\includegraphics[scale=.45]{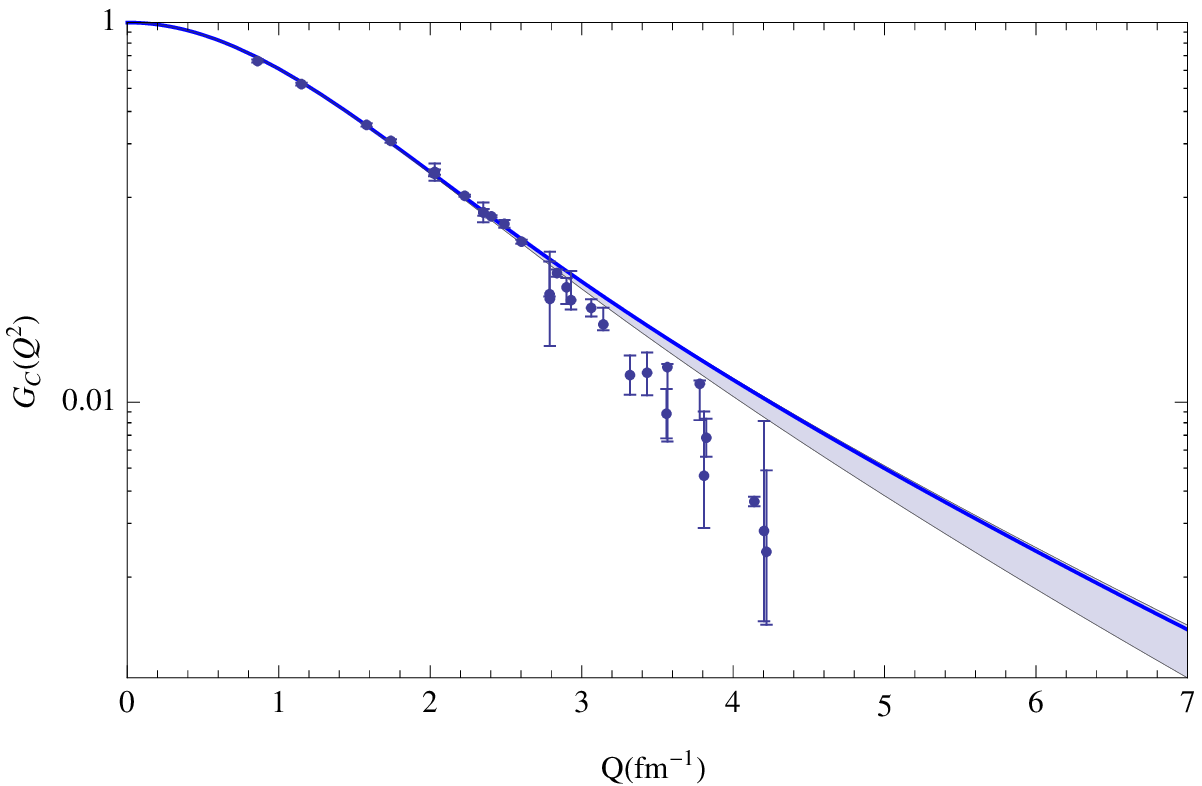}
\includegraphics[scale=.45]{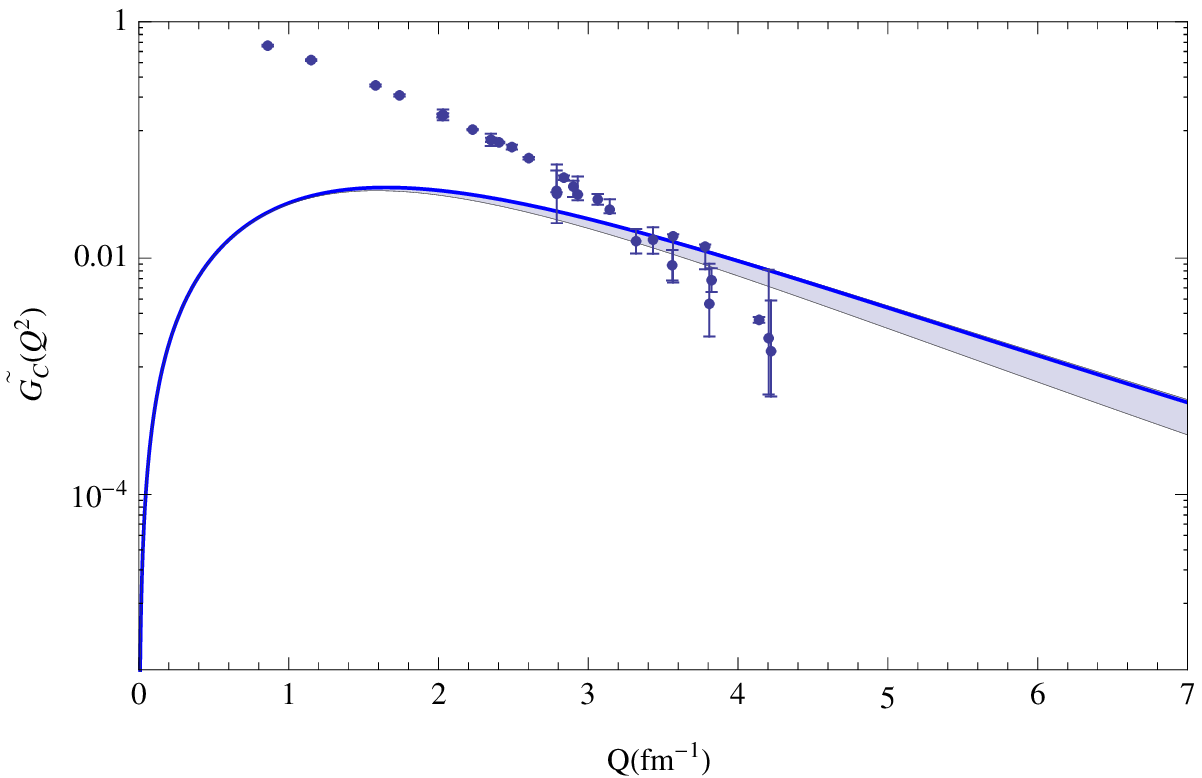}
\noindent
\caption{Results for the charge form factor $G_C(Q^2)$ with the parameters
$0 \le \alpha_2 \le 1$, $1 \le \alpha_3 \le 2$ 
(solid line corresponds to the $\alpha_2 = 0.25$ and $\alpha_3 = 1.1$) 
and $\beta = 1.2$:
exact result (left panel) and truncated result $\tilde G_C(Q^2)$
(right panel).}
\vspace*{.2cm}
\includegraphics[scale=.45]{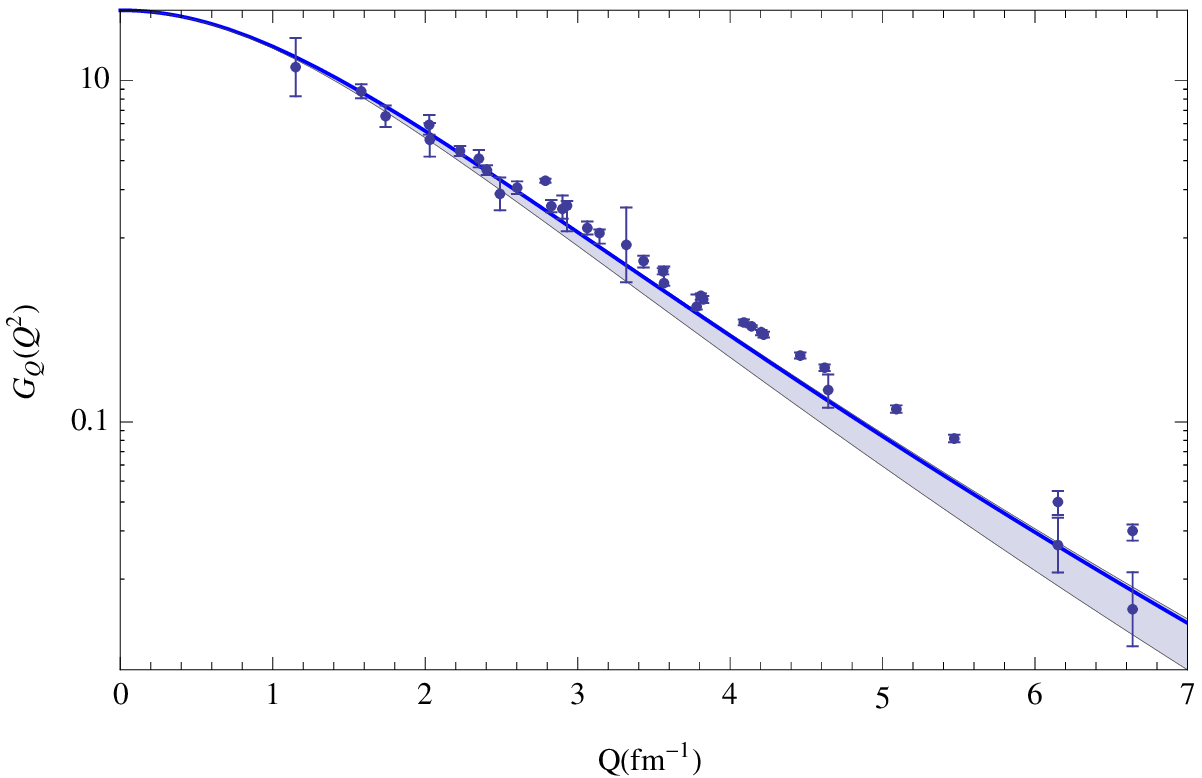}
\includegraphics[scale=.45]{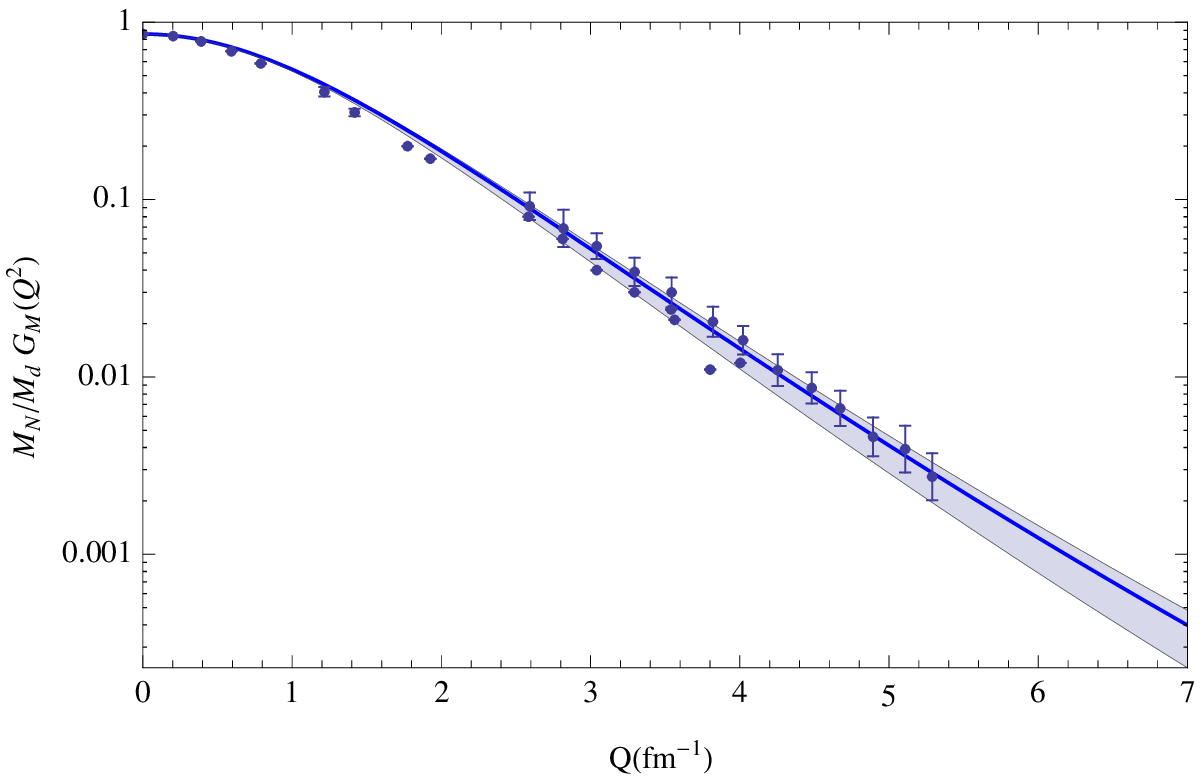}
\noindent
\caption{Quadrupole $G_Q(Q^2)$
and magnetic $(m_N/m_D) \, G_M(Q^2)$ deuteron form factors 
with the parameters 
$0 \le \alpha_2 \le 1$, $1 \le \alpha_3 \le 2$ 
(solid line corresponds to the $\alpha_2 = 0.25$ and $\alpha_3 = 1.1$) 
and $\beta = 1.2$.}
\vspace*{.2cm}
\includegraphics[scale=.45]{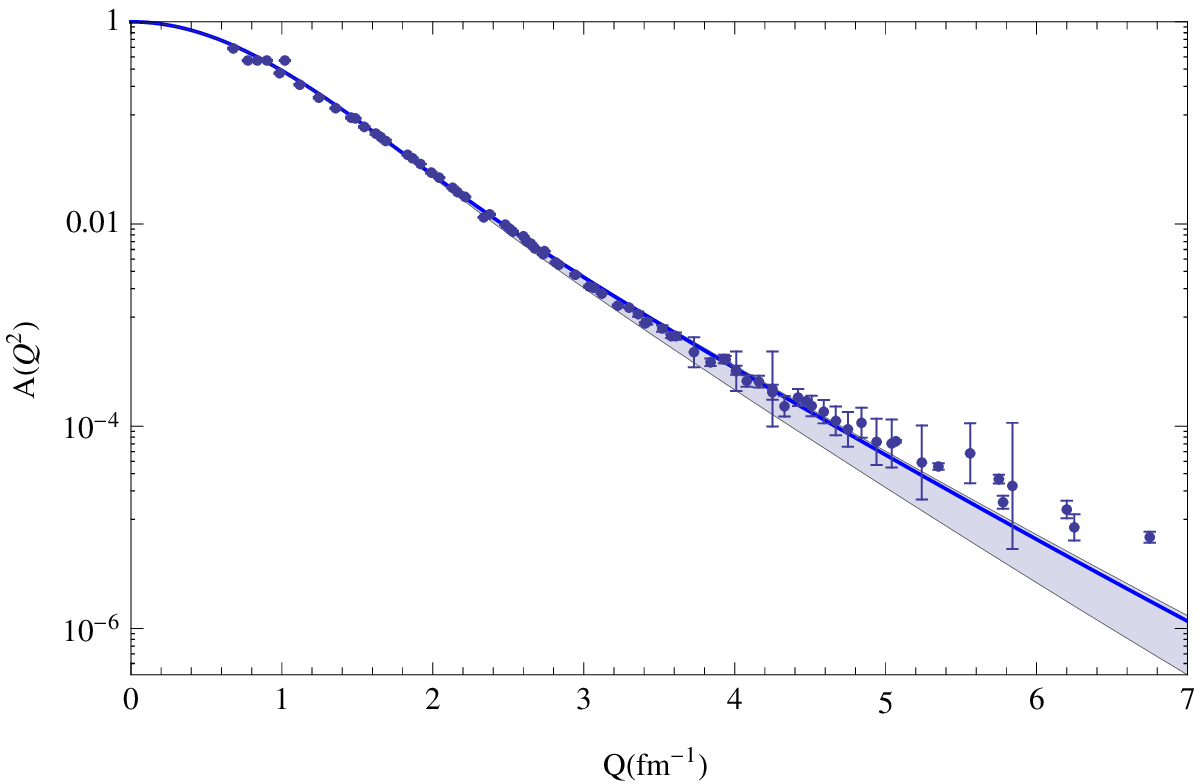}
\includegraphics[scale=.45]{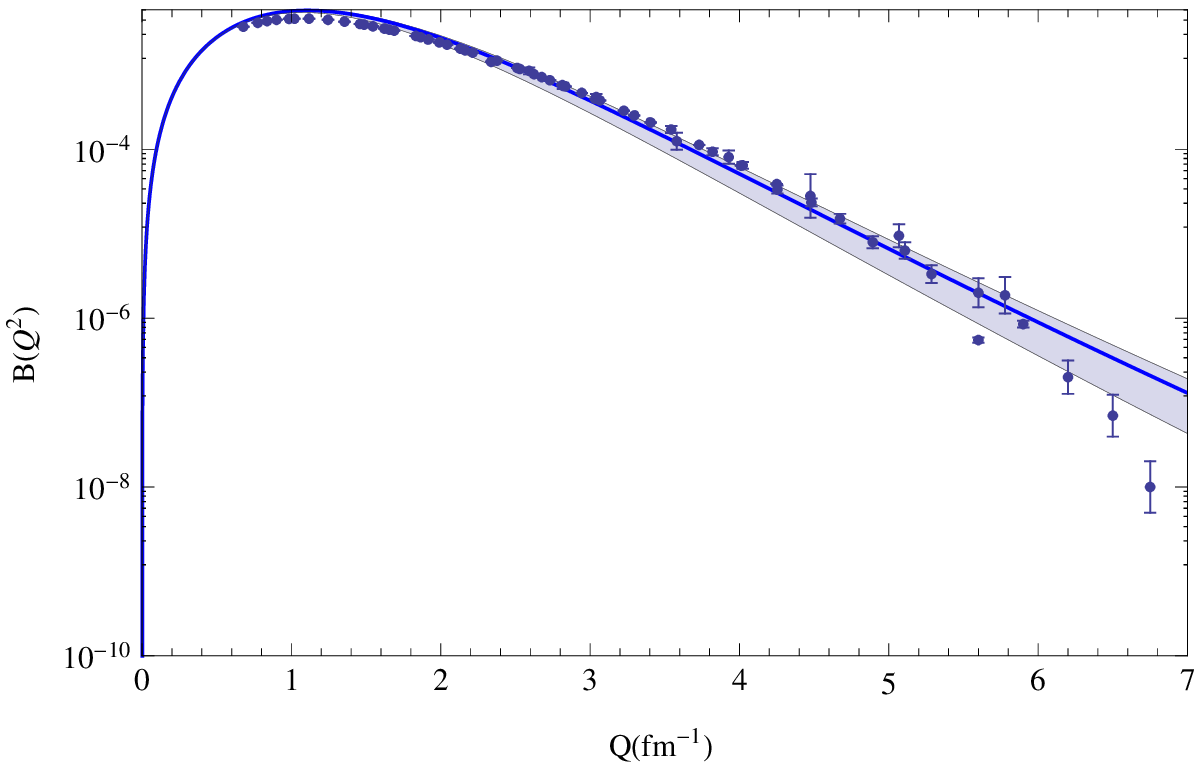}
\caption{Structure functions $A(Q^2)$ and $B(Q^2)$ with the parameters
$0 \le \alpha_2 \le 1$, $1 \le \alpha_3 \le 2$ 
(solid line corresponds to the $\alpha_2 = 0.25$ and $\alpha_3 = 1.1$) 
and $\beta = 1.2$.}
\end{center}
\end{figure}

\begin{acknowledgments}

This work was supported
by the German Bundesministerium f\"ur Bildung und Forschung (BMBF)
under Project 05P2015 - ALICE at High Rate (BMBF-FSP 202):
``Jet- and fragmentation processes at ALICE and the parton structure                                          
of nuclei and structure of heavy hadrons'',
by CONICYT (Chile) Research Project No. 80140097,
by FONDECYT (Chile) under Grants No. 1140390 and FB - 0821,
by Tomsk State University Competitiveness Improvement Program and
the Russian Federation program ``Nauka'' (Contract No. 0.1526.2015, 3854).

\end{acknowledgments}

\end{document}